\documentclass[sigconf]{acmart}
\AtBeginDocument{%
  \providecommand\BibTeX{{%
    \normalfont B\kern-0.5em{\scshape i\kern-0.25em b}\kern-0.8em\TeX}}}

\setcopyright{acmcopyright}
\copyrightyear{2018}
\acmYear{2018}
\acmDOI{XXXXXXX.XXXXXXX}

\acmConference[Conference acronym 'XX]{Make sure to enter the correct
  conference title from your rights confirmation emai}{June 03--05,
  2018}{Woodstock, NY}
\acmPrice{15.00}
\acmISBN{978-1-4503-XXXX-X/18/06}

\usepackage{multirow}
\usepackage{multicol}
\usepackage{subcaption}
\usepackage{scalefnt}

\usepackage{tikz}
\usetikzlibrary{chains,scopes,positioning,backgrounds,shapes,fit,shadows,calc,arrows.meta}
\tikzset{
    head/.style = {fill = orange!90!blue,
    label = center:\textsf{\Large L}},
    tail/.style = {fill = blue!70!yellow, text = black,
    label = center:\textsf{\large R}}
}
\usepackage{algorithm}
\usepackage{algorithmic}
\usepackage{xcolor}
\usepackage{graphicx}
\usepackage{pgfplots}
\usepgfplotslibrary{groupplots}
\usepackage{amsthm}

\usepackage{enumitem}

\definecolor{mercury}{RGB}{230,230,230}
\definecolor{gallery}{RGB}{240,240,240}
\definecolor{bgc}{RGB}{245,245,245}
\definecolor{tuatara}{RGB}{67, 67, 67}
\definecolor{free_speech_aquamarine}{RGB}{0, 156, 114}
\definecolor{shakespeare}{RGB}{85, 154, 193}
\definecolor{sun_shade}{RGB}{255, 144, 68}
\definecolor{flamingo}{RGB}{237, 88, 85}
\begin{document}

\title{Discrete Conditional Diffusion for Reranking in Recommendation}

\author{Xiao Lin}
\authornote{Both authors contributed equally to this research.}
\email{jackielinxiao@gmail.com}
\affiliation{%
  \institution{Kuaishou Technology}
  \city{Beijing}
  \country{China}
}

\author{Xiaokai Chen}
\authornotemark[1]
\email{chenxiaokai@kuaishou.com}
\affiliation{%
  \institution{Kuaishou Technology}
  \city{Beijing}
  \country{China}
}

\author{Chenyang Wang}
\email{wangchenyang07@kuaishou.com}
\affiliation{%
  \institution{Kuaishou Technology}
  \city{Beijing}
  \country{China}
}

\author{Hantao Shu}
\email{shuhantao@kuaishou.com}
\affiliation{%
  \institution{Kuaishou Technology}
  \city{Beijing}
  \country{China}
}

\author{Linfeng Song}
\email{songlinfeng@kuaishou.com}
\affiliation{%
  \institution{Kuaishou Technology}
  \city{Beijing}
  \country{China}
}

\author{Biao Li}
\email{libiao@kuaishou.com}
\affiliation{%
  \institution{Kuaishou Technology}
  \city{Beijing}
  \country{China}
}
%


\author{Peng Jiang}
\authornote{Corresponding Author.}
\email{jiangpeng@kuaishou.com}
\affiliation{%
  \institution{Kuaishou Technology}
  \city{Beijing}
  \country{China}}



\begin{abstract}

Reranking plays a crucial role in modern multi-stage recommender systems by rearranging the initial ranking list to model interplay between items. 
Considering the inherent challenges of reranking such as combinatorial searching space, some previous studies have adopted the \textit{evaluator-generator paradigm}, with a generator producing feasible sequences and a evaluator selecting the best one based on estimated listwise utility.
Inspired by the remarkable success of diffusion generative models, this paper explores the potential of diffusion models for generating high-quality sequences in reranking.
However, we argue that it is nontrivial to take diffusion models as the generator in the context of recommendation. 
Firstly, diffusion models primarily operate in continuous data space, differing from the discrete data space of item permutations. 
Secondly, the recommendation task is different from conventional generation tasks as the purpose of recommender systems is to fulfill user interests. 
Lastly, real-life recommender systems require efficiency, posing challenges for the inference of diffusion models.

To overcome these challenges, we propose a novel Discrete Conditional Diffusion Reranking (DCDR) framework for recommendation. 
DCDR extends traditional diffusion models by introducing a discrete forward process with tractable posteriors, which adds noise to item sequences through step-wise discrete operations (e.g., swapping). 
Additionally, DCDR incorporates a conditional reverse process that generates item sequences conditioned on expected user responses. 
For efficient and robust inference, we propose several optimizations to enable the deployment of DCDR in real-life recommender systems.
Extensive offline experiments conducted on public datasets demonstrate that DCDR outperforms state-of-the-art reranking methods. 
Furthermore, DCDR has been deployed in a real-world video app with over 300 million daily active users, significantly enhancing online recommendation quality.

\end{abstract}







\maketitle

\section{Introduction} 
Multi-stage recommender systems are widely adopted in online platforms like Youtube, Tiktok, and Kuaishou. As the final stage in recommender system, reranking stage takes top-ranking items as input and generates a reordered sequence of items for recommendation, and thereby directly affects users' experience and satisfaction~\cite{liu2022neural, pei2019personalized}.

Different from preliminary stages (e.g., matching, ranking) that produce predictions of a candidate item based on the item itself, the reranking stage further considers the listwise context (cross-item interplay)~\cite{xia2008listwise}.
It is widely acknowledged that whether a user is interested in an item is also determined by other items in the same list~\cite{pei2019personalized}.
Therefore, the key to reranking models is to model the listwise context and produce the optimal sequence.

Generative models are well-suited for the reranking task considering the exponentially large space of item permutations~\cite{huzhang2021aliexpress, feng2021grn}.
Previous studies have adopted the \textit{evaluator-generator paradigm}, with a generator to generate feasible permutations and an evaluator to evaluate the listwise utility of each permutation~\cite{wang2019sequential}.
In this paradigm, the capacity of generator is of great importance.
Despite the successes of traditional generative models like GANs and VAEs~\cite{higgins2017beta, wang2017irgan}, their limitations, such as unstable optimization and posterior collapse, hinder their application in the reranking task for recommendation. 
Recently, diffusion models~\cite{ho2020denoising, rombach2022high} achieve remarkable success in computer vision and other domains. 
Diffusion models usually involve a forward process that corrupts the input with noises in a step-wise manner; and a reverse process that iteratively generates the original input from the corrupted one with a denoising model.
In light of the success of diffusion models, this paper aims to explore the potential of diffusion models for generating high-quality sequences in reranking. 
However, we find it is nontrivial to take diffusion models as the generator due to the following challenges: 


\begin{itemize}[leftmargin=*]
    \item Firstly, most diffusion models~\cite{song2020denoising, ho2020denoising, nichol2021improved} are designed for continuous data domains, but the item permutations in recommender systems are operated in the discrete data space. The inherent discrete nature of item sequences in recommender systems brings challenges to the application of diffusion models.

    \item Secondly, the recommendation task is different from conventional generation tasks as the purpose of recommender systems is to fulfill user interests. The generated sequence is expected to achieve positive user feedback, and hence the diffusion model should be controllable in terms of user feedback.

    \item Thirdly, real-life recommender systems serve a huge number of users and the inference procedure of the reranking model is expected to be efficient. Since the generation process of diffusion models works in a step-wise manner, it poses challenges to the inference efficiency of diffusion models.
\end{itemize}


To tackle the aforementioned challenges, we propose a novel Discrete Conditional Diffusion Reranking (DCDR) framework, which extends traditional diffusion models with a discrete forward process and a conditional reverse process for sequence generation. 
Specifically, in each step of the forward process, DCDR uses a discrete operation to add noises to the input sequence.
We propose two discrete operations including permutation-level operation and token-level operation.
In the reverse process, DCDR introduces user feedback into the denoising model as conditions for generation. 
In each step, the denoising model takes conditions and the noisy sequence as input and estimates the distribution of denoised sequence in the next step. 
This enables the reverse process to generate sequences with expected feedback during inference.
To train the denoising model, we derive the formal objective function by introducing carefully designed sequence encoding and transition matrix for both discrete operations.
Moreover, for efficient and robust inference, we propose a series of techniques to enable the deployment of DCDR in real-life recommender systems.
We conduct extensive offline and online A/B experiments, the comparison between DCDR and other state-of-the-art reranking methods demonstrates the superiority of DCDR. 

DCDR actually serves as a general framework to leverage diffusion in reranking.
The discrete operation and model architecture in DCDR are not exhaustive, which can vary according to specific application scenarios.
We believe that DCDR will provide valuable insights for future investigations on diffusion-based multi-stage recommender systems.
The contributions of this paper can be summarized as follows:
\begin{itemize}[leftmargin=*]
    \item To the best of our knowledge, this is the first attempt to introduce diffusion models into the reranking stage in real-life multi-stage recommender systems.
    \item A novel Discrete Conditional Diffusion Reranking (DCDR) framework is presented. We carefully design two discrete operations as the forward process and and introduce user feedback as conditions to guide the reverse generation process. 
    \item We provide proper approaches for deploying DCDR in a popular video app Kuaishou, which serves over 300 million users daily. And online A/B experiments demonstrate the effectiveness and efficiency of DCDR.
\end{itemize}

\section{Related Work}
\label{related_work}

\subsection{Reranking in Recommendation}
Reranking in recommendation focuses on rearranging the input item sequence considering correlations between items to achieve optimal user feedback. 
Therefore, reranking models~\cite{pang2020setrank,pei2019personalized, xi2021context} take the whole list of items (listwise context) as input and generate a reordered list as output.
This is different from ranking models~\cite{liu2009learning,burges2005learning} in preliminary stages (e.g., matching, ranking) that consider a single candidate item at a time. 
Existing studies on reranking can be roughly categorized into two aspects: the first line of researches~\cite{pei2019personalized,pang2020setrank,ai2018learning} focus on modeling item relations and directly learn a single ranking function which ranks items greedily with the ranking score; the other line of works~\cite{huzhang2021aliexpress,feng2021grn} divides the reranking task into two components: sequence generation and sequence evaluation, with a generator produces feasible sequences an a evaluator selecting the best one based on estimated listwise utility.

Our work adopts the evaluator-generator paradigm and endeavors to explore the potential of diffusion models as the generator for producing high-quality item sequences, thereby enhancing the performance of reranking models.

\subsection{Diffusion Models}
Diffusion models~\cite{dhariwal2021diffusion,ho2020denoising,rombach2022high,huang2022prodiff} have achieved significant success in generation tasks of continuous data domains, such as image synthesis and audio generation. 
Some studies~\cite{austin2021structured,reid2023diffuser} attempt to apply diffusion models on tasks of discrete data domains like text generation. One line of researches~\cite{austin2021structured,hoogeboom2021autoregressive} design corruption processes and reconstruction processes on discrete data. Another line of researches~\cite{li2022diffusion,chen2022analog} attempt to apply continuous diffusion models on discrete data domains by adding noises to the embedding spaces of the data or real-number vector spaces. DiffusionLM~\cite{li2022diffusion} is one of the state-of-the-art method which generates text sequence with continuous diffusion models.
While diffusion models have achieved success, their potential for generating high-quality item sequences in recommendation remains under-explored. 
Recently, some studies have attempted to apply diffusion models on sequential recommendation~\cite{wang2023diffusion,wang2023conditional,li2023diffurec}, where the focus is to generate the next item based on user's historical interactions. 

However, it is important to note that the reranking task addressed in this paper is distinct from typical sequential recommendation.
Specifically, reranking aims to generate feasible item permutations rather than focusing on the next item embedding~\cite{li2023diffurec} or user vector~\cite{wang2023diffusion}, which poses significant challenges for the application of diffusion models in the reranking stage.


\section{Preliminary} 
\label{Diffusion_intro}

\subsection{Reranking Task}
Denote the ordered sequence from the previous stage as $\mathcal{R}=[i_1, i_2,\cdots,i_{l_s}]$, where $l_s$ is the input sequence length. 
The reranking problem aims to find a reordered item sequence $\mathcal{R}^{*}=[i_1^*, i_2^*,\cdots,i_{l_o}^*]$ with optimal user feedback (e.g., likes, clicks), where $l_o$ is the length of the final recommendation list.
In practice, $l_o$ is usually less than 10 and $l_s$ can either be equal to or larger than $l_o$.

\subsection{Diffusion Model}
\begin{figure}
    \centering
    \includegraphics[width=0.87\columnwidth]{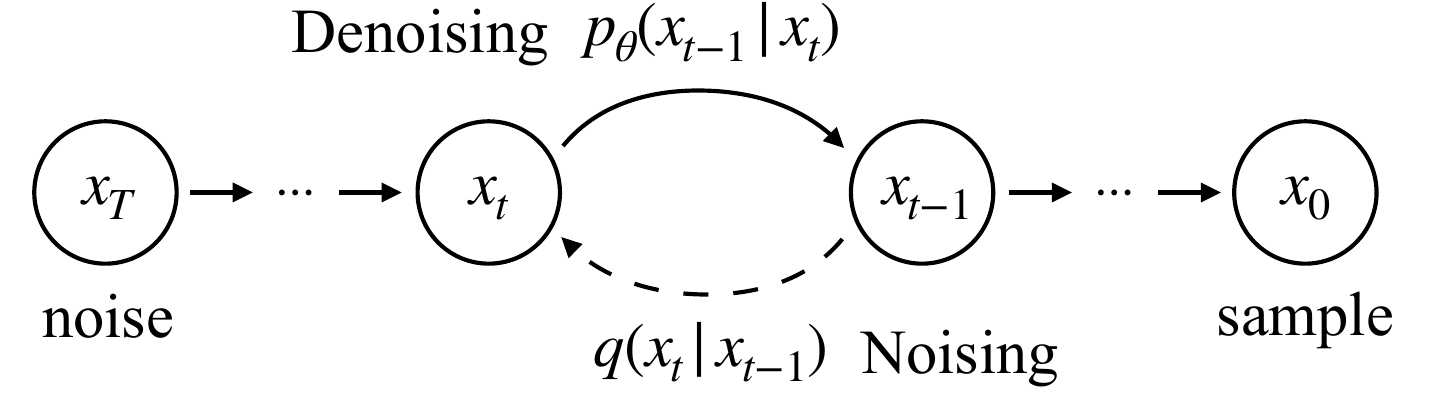}
    \caption{A graphical model illustrating the forward (noising) and reverse (denoising) processes in diffusion models. }
    \label{fig:dm_intro}
\end{figure}

Before we go deep into the details of DCDR, we first provide a brief introduction to diffusion models. 
The typical diffusion models consist of two processes: forward process and reverse process, which are illustrated in Fig.~\ref{fig:dm_intro}. During the forward phase, a sample is corrupted by adding noise in a step-wise manner, where the noise-adding process forms a Markov chain. Contrarily, the reverse phase recovers the corrupted data at each step with a denoising model.

\textit{Forward Process:}
Given a sample $\mathbf{x}_0 \sim q(\mathbf{x}_0)$, the forward Markov process $q(\mathbf{x}_{1:T}|\mathbf{x}_0)=\prod_{t}q(\mathbf{x}_t|\mathbf{x}_{t-1})$ corrupts the sample into a sequence of increasingly noisy samples: $\mathbf{x}_1,\ldots,\mathbf{x}_T$, where $t$ refers to the diffusion step. Generally the noise follows a Gaussian distribution: $q(\mathbf{x}_t|\mathbf{x}_{t-1})=\mathcal{N}(\mathbf{x}_{t-1};\sqrt{1-\beta_t}\mathbf{x}_{t-1},\beta_t\mathbf{I})$ where $\beta_t$ is the scale of the added noise at step $t$.

\textit{Reverse Process:}
The reverse Markov process attempts to recover the last-step sample with a parameterized denoising model $p_{\theta}(\mathbf{x}_{t-1}|\mathbf{x}_t)$. When the noise follows Gaussian distribution, the parameterized distribution becomes
\begin{equation}
    p_{\theta}(\mathbf{x}_{t-1}|\mathbf{x}_t)=\mathcal{N}(\mathbf{x}_{t-1}|\mu_{\theta}(\mathbf{x}_t,t),\sigma_{\theta}(\mathbf{x}_t,t)),
\end{equation}
where $\mu_{\theta}(\mathbf{x}_t,t)$ and $\sigma_{\theta}(\mathbf{x}_t,t)$ are modeled with neural networks.

\textit{Training:}
The canonical objective function \cite{ho2020denoising} is the variational lower bound: 
\begin{equation}\label{VanillaDiffuLoss}
\begin{aligned}
    \mathcal{L} = &\underbrace{D_{KL}(q(\mathbf{x}_T|\mathbf{x}_0) \lVert p(\mathbf{x}_T))}_{\mathcal{L}_T} - \underbrace{\mathbb{E}_{q(\mathbf{x}_1|\mathbf{x}_0)}[logp_{\theta}(\mathbf{x}_0|\mathbf{x}_1))]}_{\mathcal{L}_0}\\ 
    &+ \underbrace{\sum_{t=2}^TE_{q(\mathbf{x}_t|\mathbf{x}_0)}[D_{KL}(q(\mathbf{x}_{t-1}|\mathbf{x}_t,\mathbf{x}_0)\lVert p_{\theta}(\mathbf{x}_{t-1}|\mathbf{x}_t)]}_{\mathcal{L}_{t}}.
\end{aligned}
\end{equation}
Since $\mathcal{L}_T$ is a constant, it is usually removed from the loss function. $\mathcal{L}_0$ represents the reconstruction error and $\mathcal{L}_t$ represents the denoising error between denoised data at each step and the corresponding corrupted data in the forward phase.

\textit{Inference:} During inference, diffusion models start from a noisy sample $\mathbf{x}_T$ and draw denoising samples with $p_{\theta}(\mathbf{x}_{t-1}|\mathbf{x}_t)$ step by step. After $T$ steps, the generation process runs from $\mathbf{x}_T$ to $\mathbf{x}_0$ i.e. the final generated sample.


\section{Discrete Conditional Diffusion Reranking Framework}
In this section, we provide a detailed introduction to DCDR. 
First, we introduce the overall framework in Section 4.1. 
Then, we elaborate on each component from Section 4.2 to Section 4.5.
The characteristics of DCDR are discussed in Section 4.6.

\subsection{Overview of DCDR}
\begin{figure}
    \centering
    \includegraphics[width=\columnwidth]{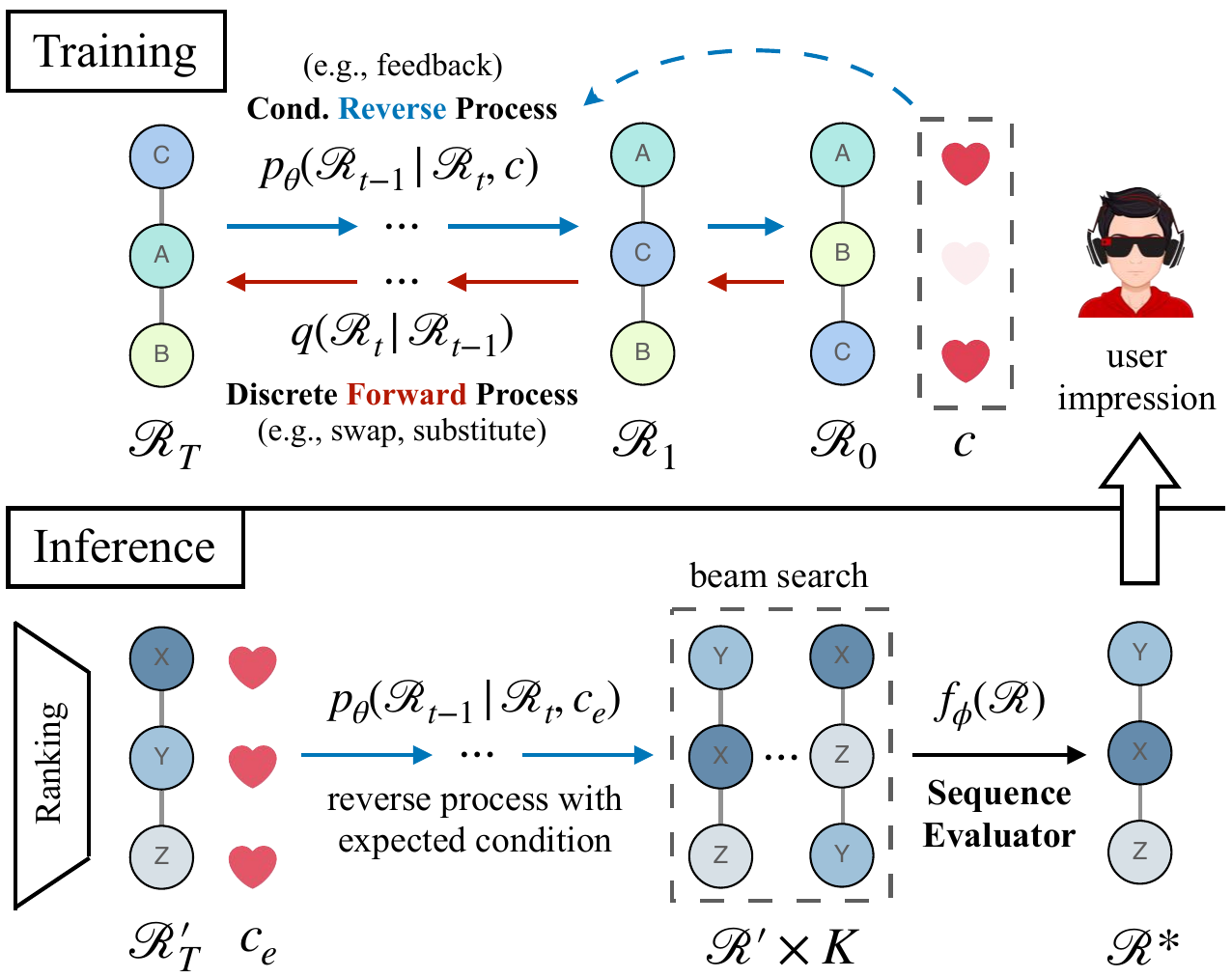}
    \caption{An illustration of the DCDR framework, which mainly consists of: 1) discrete forward process, and 2) conditional reverse process. In the forward process, two discrete operations with tractable posteriors are introduced; while in the reverse process, user feedback is introduced as the condition to control generation.}
    \label{fig:DCDR_pipeline}
\end{figure}

The framework of DCDR is illustrated in Fig.~\ref{fig:DCDR_pipeline}, which mainly consists of: 1) discrete forward process, 2) conditional reverse process. 
Specifically, the discrete forward process defines a discrete operation with tractable posteriors to add noises to the input sequence. Here we introduce permutation-level / token-level operation as an example, while other discrete operations are also feasible to be incorporated.
The conditional backward process contains a denoising model that tries to recover the input from a noisy sequence at each step. Different from traditional diffusion models, we introduce the expected feedback of the original sequence as the condition to generate the last-step sequence, which is more consistent with the recommendation task. 

During training, DCDR first adds noises to the user impression list with the discrete forward process in a step-wise manner.
Then, the denoising model in the reverse process is trained to recover the last-step sequence conditioned on user responses to the original impression list.
During inference, an initial ranked item list is fed as input, and we set an expected feedback of each item as the condition.
Then, the conditional reverse process is able to generate sequences step by step.
To further improve the generation quality, we maintain $K$ sequences with top probabilities at each step and introduce an extra sequence evaluator to select the optimal sequence for the final recommendation.

\subsection{Discrete Forward Process}

The forward process in vanilla diffusion models adds Gaussian noises to continuous signals like images according to a specified schedule. However, it is sub-optimal for sequence generation tasks as explained in aforementioned sections. Therefore, we propose to use discrete forward process for sequence generation in the reranking task.

Remember that to train a vanilla diffusion model, the variational lower bound in Eq.\eqref{VanillaDiffuLoss} involves posteriors $q(\mathbf{x}_{t-1}|\mathbf{x}_t,\mathbf{x}_0)$ in $\mathcal{L}_t$, which can be rewritten by applying Bayes' theorem:
\begin{equation}
\begin{aligned}
    q(\mathbf{x}_{t-1}|\mathbf{x}_t,\mathbf{x}_0)&=\frac{q(\mathbf{x}_t|\mathbf{x}_{t-1},\mathbf{x}_0)q(\mathbf{x}_{t-1}|\mathbf{x}_0)}{q(\mathbf{x}_{t}|\mathbf{x}_0)}\\
    &=\frac{q(\mathbf{x}_t|\mathbf{x}_{t-1})q(\mathbf{x}_{t-1}|\mathbf{x}_0)}{q(\mathbf{x}_{t}|\mathbf{x}_0)}
\end{aligned}
\end{equation}
In the continuous setting with Gaussian noises, $q(\mathbf{x}_{t}|\mathbf{x}_{t-1})$ and $q(\mathbf{x}_{t}|\mathbf{x}_0)$ are easy to compute and the posterior follows a Gaussian distribution~\cite{ho2020denoising}.
To enable discrete forward process, we need to design discrete operations that also yield tractable forward process posteriors $q(\mathcal{R}_{t-1}|\mathcal{R}_{t}, \mathcal{R}_{0})$.
Here we introduce two examples of the discrete operation, i.e., permutation-level operation and token-level operation (shown in Fig.~\ref{fig:DCDR_discrete_swap}).
Note that the DCDR framework is not limited to these two operations. Other discrete operations are also feasible to be incorporated in the future.

\subsubsection{\textbf{Permutation-level Operation}}
We first propose to encode the item sequence in the permutation space. In this setting, each sequence $\mathcal{R}_t$ is maped to an integer $v(\mathcal{R}_t)\in [0, l_o!)$ and represented as $\mathbf{s}_t\in \mathbb{R}^{1\times l_o!}$, which is a one-hot vector with length $l_{o}!$.
Here $l_o$ is the length of the final recommendation list. Then we design a simple yet effective discrete operation that forms a Markov chain in the permutation space.

For each step in the forward process, we either keep the sequence same with that in last step or randomly swap a pair of items in the sequence. This corresponds to a transition matrix $\mathbf{Q}_t\in \mathbb{R}^{l_o!\times l_o!}$ at step $t$ as follows:
\begin{equation}\label{eqn:q_matrix_perm}
[\mathbf{Q}_t]_{ij}=\left\{ \begin{aligned}
&1-\beta_t, &if \ d(v^{-1}(i), v^{-1}(j)) = 0    \\
&\beta_t/C_{l_o}^2, &if \ d(v^{-1}(i), v^{-1}(j)) = 2 \\
&0, &otherwise,
\end{aligned} \right.
\end{equation}
where $v^{-1}(\cdot)$ derives the counterpart permutation, $C_{l_o}^2=l_o(l_o-1) / 2$ is the number of possible swap candidates, $\beta_t\in[0,1]$ controls the noise strength at each step\footnote{In this work, we set $\forall t: \beta_t = \beta$ as a single hyper-parameter and leave advanced noise schedule mechanisms as future work.}. And $d(v^{-1}(i), v^{-1}(j))$ denotes the difference between two sequences. For example, given $\mathcal{R}_{t-1}=ABC$ and $\mathcal{R}_t=ACB$, we have $d(\mathcal{R}_t, \mathcal{R}_{t-1}) =2$.

\begin{figure}
    \centering
    \includegraphics[width=\columnwidth]{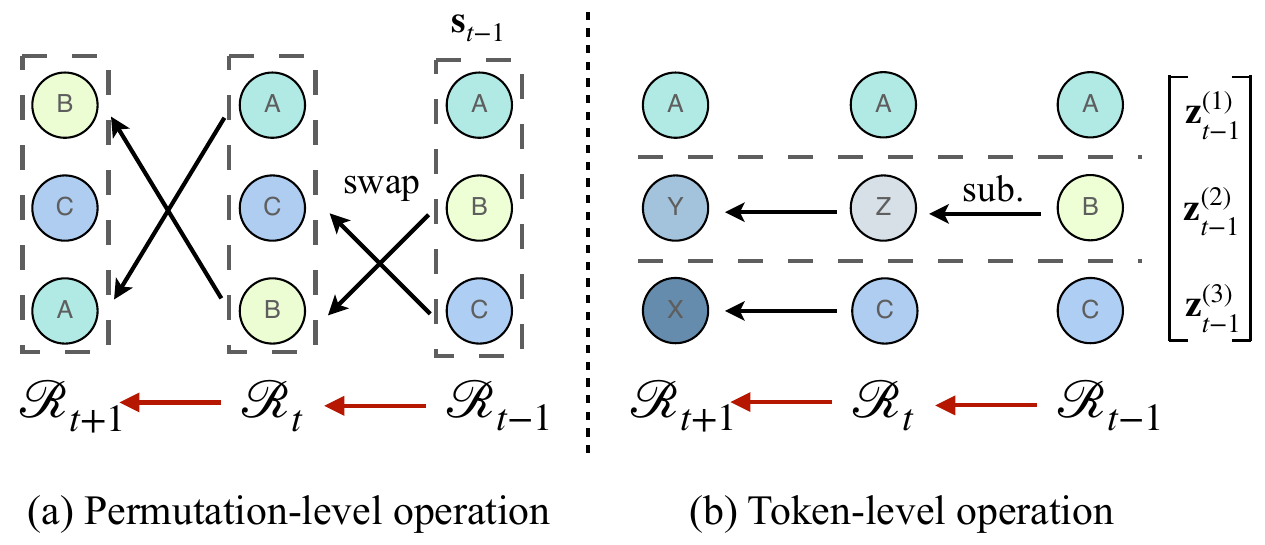}
    \caption{Illustrations of permutation-level (left) and token-level (right) operations in the discrete forward process. }
    \label{fig:DCDR_discrete_swap}
\end{figure}

We use an example in Fig.~\ref{fig:DCDR_discrete_swap}(a) to illustrate the permutation-level operation.
Notice that $d(\mathcal{R}_t , \mathcal{R}_{t-1})=d(\mathcal{R}_{t-1},\mathcal{R}_t)$, the transition matrix is a symmetric matrix. Moreover, we show that this transition matrix induces a Markov chain with a uniform stationary distribution, which means the corrupted sequence would eventually become a fully random sequence. The detailed proof can be found in the Appendix.

With the sequence encoding $\mathbf{s}_t$ and the transition matrix $\mathbf{Q}_t$, the discrete forward process at each step can be formulated as:
\begin{equation*}
    q(\mathcal{R}_t|\mathcal{R}_{t-1}) = cat(\mathbf{s}_t\text{;}~\mathbf{s}_{t-1}\mathbf{Q}_t),
\end{equation*}
where $cat(\cdot\text{;}~\mathbf{p})$ means a categorical distribution with probability $\mathbf{p}$.

After $t$ steps of noise adding, $\mathcal{R}_t$ given $\mathcal{R}_0$ can be represented as:
\begin{equation*}
    q(\mathcal{R}_t|\mathcal{R}_0) = cat(\mathbf{s}_t\text{;}~\mathbf{s}_0\overline{\mathbf{Q}}_t),
\end{equation*}
where $\overline{\mathbf{Q}}_t=\mathbf{Q}_1\mathbf{Q_2\ldots\mathbf{Q}_t}$.

As a result, we can compute the posterior $q(\mathcal{R}_{t-1}|\mathcal{R}_t,\mathcal{R}_0)$ as follows:
\begin{equation}\label{eq:permutation-level_posterior}
\begin{aligned}
    q(\mathcal{R}_{t-1}|\mathcal{R}_t,\mathcal{R}_0)
    &=\frac{q(\mathcal{R}_t|\mathcal{R}_{t-1})q(\mathcal{R}_{t-1}|\mathcal{R}_0)}{q(\mathcal{R}_{t}|\mathcal{R}_0)}\\
    &=cat(\mathbf{s}_{t-1}\text{;}~\dfrac{\mathbf{s}_t\mathbf{Q}^T_t\odot \mathbf{s}_0\overline{\mathbf{Q}}_{t-1}}{\mathbf{s}_0\overline{\mathbf{Q}}_t\mathbf{s}_t^T}).
\end{aligned}
\end{equation}
This enables us to calculate the KL-divergence in $\mathcal{L}_t$ during training.


\begin{figure*}
    \centering
    \includegraphics[width=0.95\textwidth]{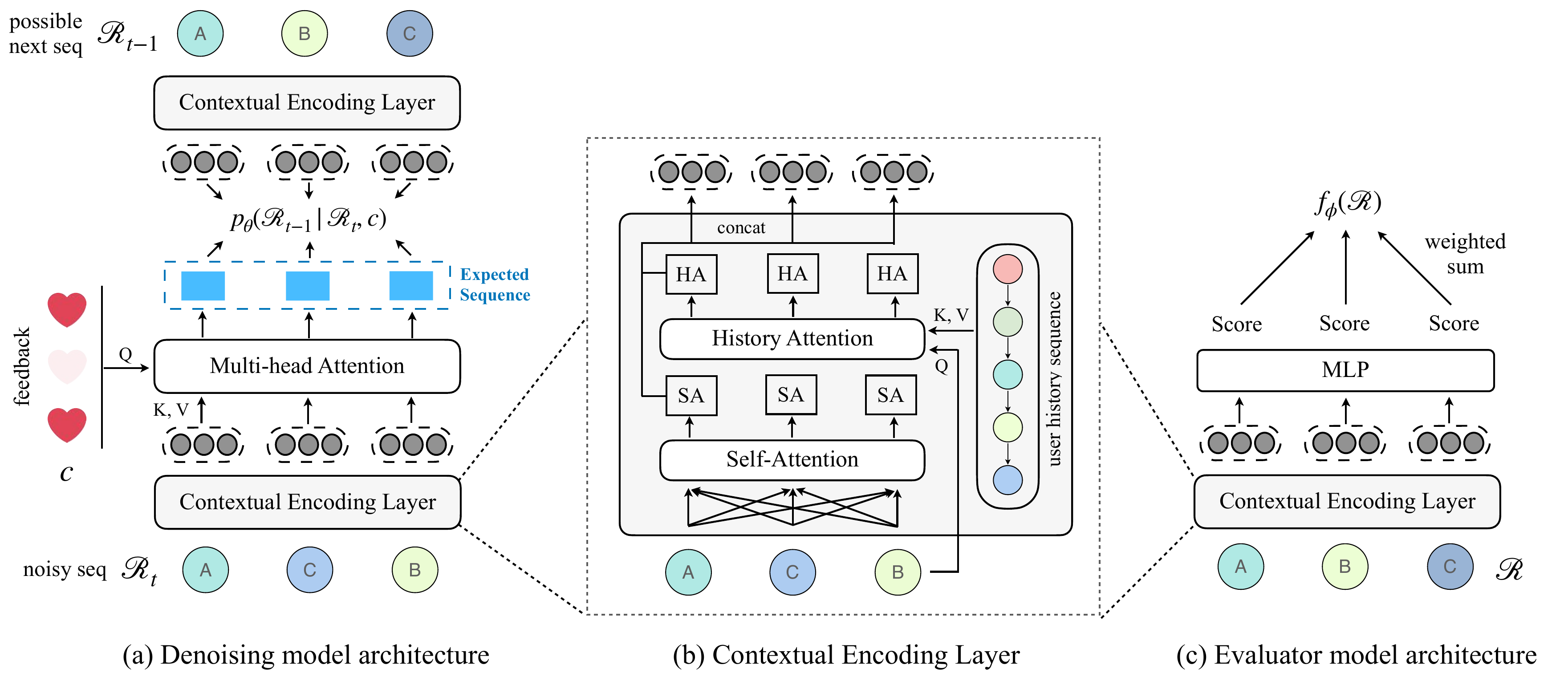}
    \caption{Illustrative instantiations of the denosing model in the conditional reverse process and the sequence evaluator model. Noted that the model architectures provided are not exhaustive, which can vary across application scenarios.}
    \label{fig:generator}
\end{figure*}

 Note that the computation of posterior at time $t$ requires a computation of $\overline{\mathbf{Q}}_t$, this can be time-consuming if we are to compute a matrix multiplication of two matrices with a size of $l_o!\times l_o!$. However, the matrix is fixed when the length of sequence $l_o$ and $\beta_t$ are determined. 
 Note that both $l_o$ and $\beta_t$ are determined before training, $\overline{\mathbf{Q}}_t$ can be computed and stored in advance. 
 Meanwhile, as $\mathbf{s}_0$ is a one-hot vector, the calculation of $\mathbf{s}_0\overline{\mathbf{Q}}_{t-1}$ is to select one row of $\overline{\mathbf{Q}}_{t-1}$ and $\mathbf{s}_0\overline{\mathbf{Q}}_{t}\mathbf{s}_t^T$ is to select an entry of the matrix. Thus the computation of posterior is essentially very efficient during training and inference.


\subsubsection{\textbf{Token-level Operation}}
Besides the permutation-level operation, we also propose a token-level operation beyond the permutation space. 
Here, we use a different way to encode an item sequence $\mathcal{R}_t$ as multiple token-level vectors $\left [\mathbf{z}_t^{(1)}\text{;}~\mathbf{z}_t^{(2)}\text{;}~\cdots\text{;}~\mathbf{z}_t^{(l_o)}\right]$, where each $\mathbf{z}_t^{(k)}\in\mathbb{R}^{1\times l_s}$ is a one-hot vector representing the item at the $k$-th position. 

For each step in the forward process, we either keep the item unchanged or substitute the item with another item from the input sequence. This corresponds with a transition matrix $\mathbf{O}_t\in \mathbb{R}^{l_s\times l_s}$ as follows:
\begin{equation}
[\mathbf{O}_{t}]_{ij} =\left\{ \begin{aligned}
&1-\beta, &if \ i=j  \\
&\frac{\beta}{l_s-1}, &otherwise
\end{aligned} \right.
\end{equation}
where $\beta$ controls the noise strength at each step. 
This forward process also induces a uniform stationary distribution, and the detailed proof can be found in the Appendix.

Then, we can formulate the discrete forward process as:
\begin{equation*}
    q(\mathcal{R}_t|\mathcal{R}_{t-1}) = \left\{cat(\mathbf{z}_t^{(k)}\text{;}~\mathbf{z}_{t-1}^{(k)}\mathbf{O}_t)\right\}_{k=1}^{l_o}.
\end{equation*}
Similarly, we can define $\overline{\mathbf{O}}_t=\mathbf{O}_1\mathbf{O_2\ldots\mathbf{O}_t}$, and the probability of $\mathcal{R}_t$ given $\mathcal{R}_0$ is represented as:
\begin{equation*}
    q(\mathcal{R}_t|\mathcal{R}_{0}) = \left\{cat(\mathbf{z}_t^{(k)}\text{;}~\mathbf{z}_{0}^{(k)}\overline{\mathbf{O}}_t)\right\}_{k=1}^{l_o}.
\end{equation*}
This leads to the posterior as follows:
\begin{equation}
    \label{eq:token-level_posterior}
    q(\mathcal{R}_{t-1}|\mathcal{R}_{0}, \mathcal{R}_{0}) = \left\{cat(\mathbf{z}_{t-1}^{(k)}\text{;}~\dfrac{\mathbf{z}_t^{(k)}\mathbf{O}^T_t\odot \mathbf{z}_0^{(k)}\overline{\mathbf{O}}_{t-1}}{\mathbf{z}_0^{(k)}\overline{\mathbf{O}}_t\mathbf{z}_t^{(k)T}})\right\}_{k=1}^{l_o}.
\end{equation}

Note that the computation of posterior also requires a calculation of $\overline{\mathbf{O}}_t$. Similarly this can be calculated and stored in advance.

\subsection{Conditional Reverse Process}
The reverse process in diffusion models attempts to recover the original sample from the noisy sample. For most diffusion models, this is parameterized as $p_{\theta}(\mathbf{x}_{t-1}|\mathbf{x}_t)$ (i.e., denoising model).
However, the recommendation task is different from conventional generation tasks as the purpose of recommender systems is to fulfill user interests.
Besides, users only response to the displayed item sequence. The utilities of other sequences are unknown. It may be problematic if we only train the denoising model to recover towards the impression list.
As a result, we expect the reverse process to be conditioned on the user feedback (e.g., like, effective view). In this way, the denoising model attempts to model $p_{\theta}({\mathcal{R}}_{t-1}|\mathcal{R}_{t},\mathbf{c})$, where $\mathbf{c}$ is a response sequence representing the expected feedback of sequence $\mathcal{R}_0$. 
During training, we can use the real feedback as condition; while at the inference stage, we can set $\mathbf{c}$ according to application scenarios (e.g., all positive feedback).

\subsubsection{\textbf{Denoising Model Architecture}}
Note that DCDR framework does not restrict the concrete architecture of the denoising model $p_{\theta}({\mathcal{R}}_{t-1}|\mathcal{R}_{t},\mathbf{c})$.
Here we give an instantiation based on attention mechanism~\cite{vaswani2017attention}, shown in Fig.~\ref{fig:generator}(a).
Each item is first mapped into an embedding vector. 
Then the sequence $\mathcal{R}_t$ goes through a contextual encoding layer, which consists of a self attention layer among items in the sequence and a history attention layer that uses items in the sequence as queries to aggregate the user history. The output of these two layers are concatenated position-wisely as the contextual-encoded sequence representations.
To introduce user feedback as condition, we map the feedback at each position as condition embeddings. Then we use these condition embeddings as queries to aggregate the contextual-encoded sequence representations of $\mathcal{R}_t$.
The outputs serve as the expected sequence representations of $\mathcal{R}_{t-1}$. 
Then the probability of drawing $\mathcal{R}_{t-1}$ is computed as the cosine similarities between the expected representations of $\mathcal{R}_{t-1}$ and the contextual-encoded sequence representations of $\mathcal{R}_{t-1}$ position-wisely.
As for the possible next sequence $\mathcal{R}_{t-1}$, it depends on the discrete operation chosen in the forward process.

\subsubsection{\textbf{Denoising Model Training}}
The training objective function of the denoising model $p_{\theta}({\mathcal{R}}_{t-1}|\mathcal{R}_{t},\mathbf{c})$ in the reverse process is the typical training loss of diffusion models:
\begin{equation}
    \mathcal{L}_t=E_{q(\mathcal{R}_t|\mathcal{R}_0)}\left[D_{KL}(q(\mathcal{R}_{t-1}|\mathcal{R}_t,\mathcal{R}_0)\lVert p_{\theta}({\mathcal{R}}_{t-1}|\mathcal{R}_t,\mathbf{c})\right].
\end{equation}
During the training stage, we randomly sample a time step $t\in \{1,\ldots, T\}$ and get the noisy sequence $\mathcal{R}_t$ via the discrete forward process (i.e. sample based on $q(\mathcal{R}_t|\mathcal{R}_0)$). Then the denoising model is optimized with $\mathcal{L}_t$ accordingly.
The training algorithm is presented in Alg.\ref{Alg:DCDRTrain}.

\begin{algorithm}[t!]
\caption{Training Algorithm for DCDR}
\label{Alg:DCDRTrain}
    \begin{algorithmic}[1]
\REQUIRE
$\mathcal{R}_0$: displayed item sequence;
$\mathbf{c}$: user feedback of $\mathcal{R}_0$;
$q(\mathcal{R}_t|\mathcal{R}_{t-1})$: discrete forward process;
$p_\theta(\mathcal{R}_{t-1}|\mathcal{R}_t,\mathbf{c})$: conditional denoising model parameterized by $\theta$;
$T$: maximal step
\REPEAT
\STATE Sample $t\sim \mathcal{U}(1,T)$;
\STATE Compute $\mathcal{R}_t$ given $\mathcal{R}_0$ via discrete forward process;
\STATE Compute $q(\mathcal{R}_{t-1}|\mathcal{R}_t,\mathcal{R}_0)$ according to Eq.\eqref{eq:permutation-level_posterior} or Eq.\eqref{eq:token-level_posterior};
\STATE Compute $p_\theta(\mathcal{R}_{t-1}|\mathcal{R}_t,\mathbf{c})$;
\STATE Compute $\mathcal{L}_t$ and update $\theta$ with $\nabla_{\theta}\mathcal{L}_t$;
\UNTIL{converged}
    \end{algorithmic}
\end{algorithm}

\subsection{Inference of DCDR}
The inference procudure in vanilla diffusion models starts with a pure Gaussian noise $\mathbf{x}_T$ and samples $\mathbf{x}_{t-1}$ step by step. To accommodate the recomendation reranking scenario, we make some changes to enable the deployment of DCDR in real-life recommender systems.

\begin{itemize}[leftmargin=*]
    \item \textbf{Condition setting}: To fullfill user interests as much as possible, we set the expected condition $\mathbf{c}_e$ as a sequence with all positive feedback during inference.
    
    \item \textbf{Starting sequence}: In traditional diffusion models, the inference starts with a pure Gaussian noise. However, the pure Gaussian noise may eliminate the important information like user preferences thus hurting recommendation quality. More importantly, starting from pure noise demands far more steps for high-quality generation. Therefore, we use the ordered sequence from the previous stage as the starting sequence for generation.
    
    \item \textbf{Beam search}:
    To improve the robustness of the sequence generation process, we adopt beam search to generate a specific number of sequences and further adopt a sequence evaluator model (detailed in Section 4.5) to select the final recommended sequence. 
    Starting from the first noisy sequence, we use the diffusion model to generate $K$ sequences for each candidate and keep $K$ sequences with top-probabilities at each step\footnote{Notice that when adopted token-level operation, the reverse process is possible to generate sequences containing duplicated items. We manually filter out such sequences and only retain sequences without duplicated items at each step.}.
    
    \item \textbf{Early stop}:
    Despite the inference optimizations like starting sequence and beam search, the multi-step reverse process may still be costly in time. Therefore we further introduce an early-stop mechanism into the inference procedure. 
    For each step, we compute the likelihood that generated sequences match the expected conditions. With the diffusion steps increasing, the likelihood is expected to increase accordingly. And the diffusion steps would terminate when the likelihood stops increasing or the increase becomes quite marginal.
    
\end{itemize}

The detailed inference algorithm of DCDR is described in Alg.\ref{Alg:DCDRInf}.

\begin{algorithm}[t!]
\caption{Inference Algorithm for DCDR}
\label{Alg:DCDRInf}
    \begin{algorithmic}[1]
\REQUIRE
$\mathcal{R}_T$: initial item sequence;
$\mathbf{c}_e$: expected user feedback;
$p_{\theta}(\mathcal{R}_{t-1}|\mathcal{R}_t,\mathbf{c})$: conditional denoising model; 
$T$: maximal step; 
$K$: number of candidate sequences;
$S_c$: sequence candidates set; 
$f_{\phi}(\mathcal{R})$: sequence evaluator model
\STATE $S_c=\{\mathcal{R}_T\}$;
\FOR{$t=\{T,T-1,\ldots,1\}$}
    \FOR{$\mathcal{R}_t \in S_c$}
        \STATE Compute $p_\theta(\mathcal{R}_{t-1}|\mathcal{R}_{t},\mathbf{c}_e)$ given $\mathcal{R}_t$ and $\mathbf{c}_e$;
        \STATE Sample $K$ item sequences according to $p_\theta(\mathcal{R}_{t-1}|\mathcal{R}_{t},\mathbf{c}_e)$ with highest probabilities;
    \ENDFOR
    \STATE Merge the sampled item sequences into $S_c$ and keep $K$ item sequences with the overall highest probabilities;
    \IF{meet early stop criterion}
        \STATE break;
    \ENDIF
\ENDFOR
\STATE Select the final sequence with $f_{\phi}(\mathcal{R})$ for recommendation;
\end{algorithmic}
\end{algorithm}

\subsection{Sequence Evaluator}
The sequence evaluator model $f_\phi(\mathcal{R})$ mainly focuses on estimating the overall utility of a given sequence.
Note that many existing methods for the sequence evaluator~\cite{feng2021revisit, wei2020generator} can be incorporated in our DCDR framework.
To center the contribution of this paper to the overall framework, we adopt an intuitive design of the sequence evaluator.
The architecture of the evaluator model is depicited in Fig.~\ref{fig:generator}(c).
Specifically, the input sequence is encoded by the contextual encoding layer as that in the denoising model.
Then, the representation at each position is fed into a MLP to predict the score of a given feedback label.
The overall utility of the sequence is measured by the rank-weighted sum of scores at each position.

Notice that the feedback label may vary across different recommendation tasks, such as clicks and purchases in e-commerce services, likes and forwards in online social media. 
Without loss of generality, we utilize the same feedback label as that in the conditional reverse process, and the objective function is a binary cross-entropy loss.

\subsection{Discussion}
The proposed DCDR provides a general and flexible framework to utilize diffusion models in recommendation.
Various discrete operations applied in the forward process yield different diffusion models.
For instance, the proposed permutation-level operation is well-suited for scenarios where the set of items to be displayed has been fixed (i.e., $l_s = l_o$).
The diffusion model learns how to generate the optimal permutation by iteratively swaping items conditioned on the expected feedback.
Conversely, the token-level operation is suitable when the final displayed item list is a subset of the initial sequence (i.e., $l_s > l_o$).
The diffusion model learns how to generate the best sequence by step-wise substituting each item with a candidate item.
Researchers can also develop other discrete operations tailored to specific application scenarios.
Moreover, the architectures of the denoising model and the sequence evaluator model are also flexible to incorporate specific contextual features.
Consequently, we believe that DCDR will provide valuable insights for future investigations on diffusion-based multi-stage recommender systems.

\section{Experiments} 
\label{experiments}
We conduct extensive experiments in both offline and online environments to demonstrate the effectiveness of DCDR. Three research questions are investigated in the experiments:
\begin{itemize}[leftmargin=*]
    \item \textbf{RQ1}: How does DCDR perform in comparison with state-of-the-art methods for reranking in terms of recommendation accuracy and generation quality?
    \item \textbf{RQ2}: How do different settings of DCDR affect the performance?
    \item \textbf{RQ3}: How does DCDR perform in real-life recommender systems?
\end{itemize}

\subsection{Experiment Setup}
\subsubsection{\textbf{Dataset}}
For the consistency of dataset and the problem setup, we expect that each sample of the dataset is a real item sequence displayed in a complete session rather than a list constructed manually. Therefore we collect two datasets for offline experiments:
\begin{itemize}[leftmargin=*]
    \item \textbf{Avito}\footnote{https://www.kaggle.com/c/avito-context-ad-clicks/data.}: this is a public dataset consisting of user search logs. Each sample corresponds to a search page with multiple ads and thus is a real impressed list with feedback. The data contains over 53M lists with 1.3M users and 36M ads. The impressions from first 21 days are used as training and the impressions in last 7 days are used as testing. And each list has a length of 5.
    \item \textbf{VideoRerank}: this dataset is collected from Kuaishou, a popular short-video App with over 300 million daily active users. For the consistency of dataset and the problem setup, each sample in the dataset is also a real item sequence displayed to a certain user in a complete session. The dataset contains $100,102$ users, $1,243,877$ items and $871,834$ lists where each list contains 6 items.
\end{itemize}


\subsubsection{\textbf{Baselines}}
We compare the proposed DCDR with serveral state-of-the-art reranking methods and a modified discrete diffusion method for text generation.
\begin{itemize}[leftmargin=*]
    \item \textbf{PRM}~\cite{pei2019personalized}: PRM is one of the state-of-the-art models for reranking tasks, which uses self attention to capture the relations between items. And it has been used for reranking tasks in Taobao recommender systems.
    \item \textbf{DLCM}~\cite{ai2018learning}: DLCM adopts gated recurrent units to model the cross-item relations sequentially. Meanwhile DLCM is optimized with an attention-based loss function, which also contributes to its predictive accuracy.
    \item \textbf{SetRank}~\cite{pang2020setrank}: SetRank tries to learn a permutation-invariant model for reranking by removing positional encodings and generates the sequence with greedy selection of the items.
    \item \textbf{EGRerank} \cite{huzhang2021aliexpress}: EGRerank consists of a sequence generator and a sequence discriminator. And the generator is optimized with reinforcement learning to maximize the expected user feedbacks under the guidance of evaluator. It is worth noticing that EGRerank has been used for reranking tasks in AliExpress.
    \item \textbf{DiffusionLM-R}~\cite{li2022diffusion}: DiffusionLM is originally proposed for generating text sequences, which is similar to the reranking task. We treat items as words and modify DiffusionLM to take the same conditions with DCDR as inputs for controllable generation.
\end{itemize}

\subsubsection{\textbf{Implementation Details}}
For different datasets, we use different user feedback as condition in the reverse process of DCDR.
For Avito, we use the click behavior as feedback.
For VideoRerank, we set the feedback condition as a binary value indicating whether a video has been completed watched.
The settings for hyper-parameters of DCDR can be found in Section 5.2.2.
As for other baselines, we carefully tune corresponding hyper-parameters to achieve the best performance.

\subsubsection{\textbf{Metrics}}
We use two widely-adopted metrics, namely AUC and NDCG, to evaluate the performance of different methods in the offline experiments.

\subsection{Offline Experiment Results}
For all offline experiments, we first pretrain a ranking model with Lambdamart and use the ranked list as the initial sequence for the reranking task.
For our DCDR, the variant using permutation-level operation is denoted as \textbf{DCDR-P}, while the variant using token-level operation is denoted as \textbf{DCDR-T}.

\subsubsection{\textbf{Performance Comparison (RQ1)}}
The performances of different approaches are listed in Table~\ref{tab:offline_exp_comparison}. Notice that DCDR achieves the best performances over other approaches, this verifies the effectiveness of DCDR in item sequence generation quality. 
Moreover the comparison between DCDR and DiffusionLM-R indicates that the generation quality by discrete diffusion models in DCDR outperforms DiffusionLM-R significantly. This verifies the benefits of discrete diffusion models for discrete data domains.

    
\begin{table}
  \caption{Comparison between DCDR and other baselines. *~means significant improvements with $p<0.05$.}
  \label{tab:offline_exp_comparison}
  \begin{tabular}{ccccc}
    \toprule
    \multirow{2}{*}{Methods} & \multicolumn{2}{c}{Avito} & \multicolumn{2}{c}{VideoRerank}\\
    \cmidrule(lr){2-3} \cmidrule(lr){4-5}
    \multirow{2}{*}{} & AUC & NDCG@3 & AUC & NDCG@3\\
    \midrule
    PRM & 0.8651 & 0.3559 & 0.6174 & 0.6317 \\
    EGRerank & 0.8724 & 0.3570 & 0.6171 & 0.6325 \\
    SetRank & 0.8663 & 0.3564 & 0.6037 & 0.6235 \\
    DLCM & 0.8982 & 0.3578 & 0.6054 & 0.6251 \\
    DiffusionLM-R & 0.9031 & 0.3766 & 0.6302 & 0.6422 \\
    \midrule
    DCDR-T  & $\mathbf{0.9120}^{*}$ & $\mathbf{0.3818}^{*}$  & $\mathbf{0.6340}^{*}$ & $\mathbf{0.6522}^{*}$ \\
    DCDR-P  & $\mathbf{0.9172}^{*}$ & $\mathbf{0.3901}^{*}$  & $\mathbf{0.6361}^{*}$ & $\mathbf{0.6576}^{*}$ \\
  \bottomrule
\end{tabular}
\end{table}


\subsubsection{\textbf{Analysis of DCDR (RQ2)}}
In this section, we provide a detailed analysis of DCDR from multiple aspects.

\pgfplotsset{
axis background/.style={fill=gallery},
grid=both,
  xtick pos=left,
  ytick pos=left,
  tick style={
    major grid style={style=white,line width=1pt},
    minor grid style=bgc,
    draw=none
    },
  minor tick num=1,
  ymajorgrids,
  major grid style={draw=white},
  y axis line style={opacity=0},
  tickwidth=0pt,
}


\begin{figure*}[!htbp]
\centering
    \begin{tikzpicture}[scale=0.55]
  \begin{groupplot}[
      group style={group size=3 by 1,
          horizontal sep = 100pt
          }, 
        ymajorgrids,
        major grid style={draw=white},
        y axis line style={opacity=0},
        tickwidth=0pt,
        yticklabel style={
        /pgf/number format/fixed,
        /pgf/number format/precision=5
        },
        scaled y ticks=false,
        every axis title/.append style={at={(0.1,0.8)},font=\bfseries}
      ]
    \nextgroupplot[
   xlabel= Beam Size $K$,
        ylabel=NDCG@3,
        legend pos=south east,
        ymin = 0.3700, ymax = 0.3920
    ]
    \addplot[thick,color=black,mark= otimes] coordinates {
          (2,0.3806)
          (4,0.3859)
          (6,0.3901)
          (8,0.3880)
          (10,0.3844)
          
        };\addlegendentry{DCDR-P}
    \addplot[thick,color=flamingo,mark=square] coordinates {
          (2,0.3778)
          (4,0.3804)
          (6,0.3818)
          (8,0.3825)
          (10,0.3748)
          
        };\addlegendentry{DCDR-T}

        \nextgroupplot[
   xlabel= Reverse Steps $T$,
        ylabel=NDCG@3,
        legend pos=south east,
        ymin = 0.3780, ymax = 0.3920
    ]
    \addplot[thick,color=black,mark= otimes] coordinates {
          (1,0.3792)
          (2,0.3861)
          (3,0.3901)
          (4,0.3890)
          (5,0.3844)
          
        };\addlegendentry{DCDR-P}
    \addplot[thick,color=flamingo,mark=square] coordinates {
          (1,0.3785)
          (2,0.3801)
          (3,0.3818)
          (4,0.3852)
          (5,0.3828)
          
        };\addlegendentry{DCDR-T}

            \nextgroupplot[
   xlabel= Noise Scale $\beta$,
        ylabel=NDCG@3,
        legend pos=south east,
        ymin = 0.3700, ymax = 0.3920
    ]
    \addplot[thick,color=black,mark= otimes] coordinates {
          (0.1,0.3847)
          (0.2,0.3901)
          (0.3,0.3903)
          (0.4,0.3821)
          (0.5,0.3813)
          
        };\addlegendentry{DCDR-P}
    \addplot[thick,color=flamingo,mark=square] coordinates {
          (0.1,0.3759)
          (0.2,0.3818)
          (0.3,0.3812)
          (0.4,0.3791)
          (0.5,0.3796)
        };\addlegendentry{DCDR-T}

  \end{groupplot}
\end{tikzpicture}

    \caption{The performances of DCDR-P and DCDR-T on Avito dataset with different hyper-parameter settings, including beam size $K$ in reverse process for inference, the number of diffusion steps $T$ and the noise scale $\beta$ in forward process.}
    \label{fig:exp_settings}
\end{figure*}
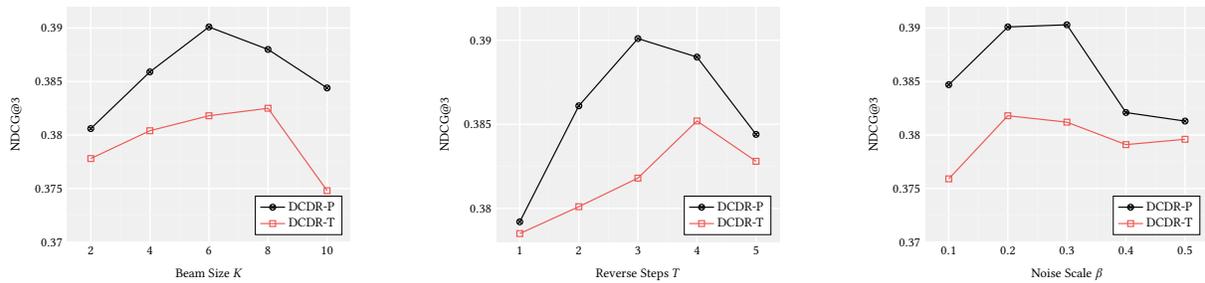

  

\textit{Beam size $K$:} We alter the beam size in the reverse process for inference and the results are presented in Fig. \ref{fig:exp_settings}. Note that the result is best when beam size is set to 6, this makes sense since a proper number of beam size improves the recommendation robustness without involving too many sequences to evaluate.

\textit{Reverse steps $T$:} We alter the number of reverse steps from 1 to 5 and list the performances in Fig.~\ref{fig:exp_settings}.
As shown in the figure, the performance gets improved when the number of diffusion steps increases. However the improvement  becomes marginal when the number of steps reaches a certain value. This coincides with the intuition of adopting early-stop for efficiency.



\textit{Noise scale $\beta$:} We alter the the swapping probability from $0.1$ to $0.5$ during training and present the results in Fig. \ref{fig:exp_settings}. The result indicates that too much noise may add difficulty to the learning process and a proper amount of noise leads to satisfactory performances.

\subsection{Online Experiment Results (RQ3)}
We conduct online A/B experiments on KuaiShou APP. 

\subsubsection{\textbf{Experiment Setup}}
In online A/B experiments, the traffic of the whole app is split into ten buckets uniformly. $20\%$ of traffic is assigned to the online baseline PRM while another $10\%$ is assigned to DCDR-P. As revealed in \cite{zhan2022deconfounding}, Kuaishou serves over 320 million users daily, and the results collected from $10\%$ of traffic for a whole week is very convincing.
The initial video sequence comes from the early stage of the recommender system, which greedily ranks the items with point-wise ranking scores. And we directly use the initial sequence as the noisy sequence $\mathcal{R}_T$ for generation.
To enable the controllable video sequence generation, we set two conditions for diffusion models: first, we expect the users to finish watching each video in the sequence; second, we expect the positive feedback from users (for example like the video)

\subsubsection{\textbf{Experiment Results}}
The experiments have been launched on the system for a week, and the results are listed in Table \ref{tab:online_exp_comparison}. 
\begin{table}
  \caption{Online experiment results. All values are the relative improvements of DCDR-P over the baseline PRM.
  For Online A/B tests in Kuaishou, the improvement of over $0.5\%$ in positive interactions (Like, Follow, Collect, Download) and $0.3\%$ in Views is very significant.} 
  \label{tab:online_exp_comparison}
  \begin{tabular}{ccccc}
    \toprule
    Views &  Likes & Follows & Collects & Downloads \\
    \midrule
    $+0.341\%^*$ & $+0.884\%^*$& $+1.100\%^*$ & $+1.299\%^*$ & $+1.358\%^*$  \\
  \bottomrule
\end{tabular}
\end{table}
The metrics for online experiments include views, likes, follows (subscriptions of authors), collects (adding videos to collections) and downloads. Notice that DCDR-P outperforms the baseline in all the related metrics by a large margin. This verifies the quality of recommended video sequence of DCDR-P.

\subsubsection{\textbf{Efficiency of DCDR}}
As diffusion models generates the samples in a step-wise manner, the cost of computation and latency become a challenge for the deployment in real-life recommender systems. The computation costs and service latency are listed in Table \ref{tab:online_exp_comparison_efficiency}.
The step-wise generation causes inevitable latency costs of the recommendation service. However, the cost is acceptable in the system since it does not jeopardize the user experience.




\begin{table}
  \caption{The additional cost of computation to the system and the additonal latency of reranking service. The computation is measured by CPU time. Avg (Comp)/Max (Comp) measure the average/maximum computation costs during the launch time of experiments respectively.} 
  \label{tab:online_exp_comparison_efficiency}
  \begin{tabular}{cccc}
    \toprule
    Avg (Comp) & Max (Comp) & Avg (Latency) & Max (Latency) \\
    \midrule
    $+0.055\%$& $+0.198\%$ & $+0.610\%$ & $+0.696\%$  \\
  \bottomrule
\end{tabular}
\end{table}


\section{Conclusion}
In this paper, we make the first attempt to enhance the reranking stage in recommendation by leveraging diffusion models, which faces many challenges such as discrete data space, user feedback incorporation, and efficiency requirements.
To address these challenges, we propose a novel framework called Discrete Conditional Diffusion Reranking (DCDR), involving a discrete forward process with tractable posteriors and a conditional reverse process that incorporates user feedback.
Moreover, we propose several optimizations for efficient and robust inference of DCDR, enabling its deployment in a real-life recommender system with over $300$ million daily active users.
Extensive offline evaluations and online experiments demonstrate the effectiveness of DCDR.
The proposed DCDR also serves as a general framework.
Various discrete operations and contextual features are flexible to be incorporated to suit different application scenarios.
We believe DCDR will provide valuable insights for future investigations on diffusion-based multi-stage recommender systems.
In the future, we plan to study the impact of noise schedule and explore methods to enhance the efficiency and controllability of the generation process.

\section{Appendix}
\begin{lemma}
    Let $\mathbf{P}$ be the transition matrix of a Markov chain.
    If $\mathbf{P}$ is a doubly-stochastic matrix, then the Markov chain defined by $\mathbf{P}$ has a uniform stationary distribution.
\end{lemma}
\begin{proof}
    Given a transition matrix $\mathbf{P}$, there exists a collection of eigenvalues and eigenvectors.
    As $\mathbf{P}$ is doubly-stochastic (every row sums to 1 and every column sums to 1), it is easy to verify that 1 is an eigenvalue of $\mathbf{P}$, with a corresponding eigenvector $\mathbf{e} / n$ ($1\cdot \mathbf{e} / n = \mathbf{P}\cdot \mathbf{e} / n = \mathbf{e} / n$).
    Therefore, we have $\mathbf{\pi} = \mathbf{e} / n$ and $\mathbf{\pi} = \mathbf{\pi}\mathbf{P}$, indicating that uniform distribution is a stationary distribution of the Markov chain.
\end{proof}

\begin{lemma}
    A Markov chain is ergodic if and only if the process satisfies 1) connectivity: $\forall i,j: \mathbf{P}^k(i,j) > 0$ for some $k$, and 2) aperiodicity: $\forall i: gcd\{k: \mathbf{P}^k(i,j)>0\} = 1$.
    And any ergodic Markov chain has a unique stationary distribution.
\end{lemma}
\begin{proof}
    The details can be found in the reference~\cite{norris1998markov}.
\end{proof}

\begin{theorem}
    The Markov chain for the discrete forward process with permutation-level operation or token-level operation has a unique uniform stationary distribution.
\end{theorem}
\begin{proof}
    First, it is easy to verify that both transition matrices are doubly stochastic: $\forall j: \sum_i[\mathbf{Q}]_{ij} = 1$ and $\forall i: \sum_j[\mathbf{O}]_{ij} = 1$.
    Therefore, uniform distribution is a stationary distribution of both processes according to Lemma 7.1.

    Meanwhile, we can show that both Markov chains are ergodic.
    For the permutation-level operation, any permutation can be achieved through finite steps of swaps. 
    For the token-level operation, it is easy to verify that each item has a chance to appear at each position.
    Therefore, both forward processes satisfy the connectivity condition.
    Besides, notice that $[\mathbf{Q}]_{ii} > 0$ and $[\mathbf{O}]_{ii} > 0$, thus the set $\{k: \mathbf{Q}^k(i,j) > 0\}$ and $\{k: \mathbf{O}^k(i,j) > 0\}$ contain 1.
    This indicates that $gcd\{k: \mathbf{Q}^k(i,j) > 0\} = 1$ and $gcd\{k: \mathbf{O}^k(i,j) > 0\} = 1$, which satisfies the aperiodicity condition.
    Therefore, both forward processes are ergodic, and hence only one stationary distribution exists according to Lemma 7.2.

    Combining the above conclusions, both processes have a unique uniform stationary distribution, which means that the sequences are almost randomly arranged after a sufficient number of steps.
\end{proof}

\balance
\bibliographystyle{ACM-Reference-Format}
\bibliography{sample-sigconf}
\end{document}